\documentclass{ws-ijmpa}

\begin{document}

%

\def\nocropmarks{\vskip5pt\phantom{cropmarks}}

\let\trimmarks\nocropmarks      

%

\markboth{Hideshi Baba et al.}
{Twist-3 effects in polarized photon structure}

%
\catchline{}{}{}
%

\setcounter{page}{1}

\title{
TWIST-3 EFFECTS IN POLARIZED PHOTON STRUCTURE
\footnote{Presented by T. Uematsu at the 3rd Circum-Pan-Pacific
Symposium on \lq\lq High Energy Spin Physics\rq\rq, Beijing, 
October 8-13, 2001.\ KUCP-204, YNU-HEPTh-02-104, to 
appear in the Proceedings.
}
}

\author{\footnotesize HIDESHI BABA\footnote{
e-mail address: baba@phys.h.kyoto-u.ac.jp}}

\address{Graduate School of Human and Environmental Studies,
Kyoto University \\ Kyoto, 606-8501, Japan}

\author{KEN SASAKI\footnote{
e-mail address: sasaki@cnb.phys.ynu.ac.jp}}

\address{Dept. of Physics, Faculty of Engineering, Yokohama National
University\\
Yokohama, 240-8501, Japan}
\author{TSUNEO UEMATSU\footnote{e-mail address:
uematsu@phys.h.kyoto-u.ac.jp}}

\address{Dept. of Fundamental Sciences, FIHS, Kyoto University\\
Kyoto, 606-8501, Japan}

\maketitle


\begin{abstract}
The polarized photon structure is described by two spin structure functions 
$g_1^\gamma$ and $g_2^\gamma$ which can be studied in the future polarized 
ep or e$^+$e$^-$ colliders. Here we investigate the QCD twist-3 effects in 
$g_2^\gamma$ to the leading order in QCD. 
\end{abstract}

\section{Introduction}	
In recent years there has been growing interest in the 
polarized photon structure functions. Especially the first moment of
a photon structure function $g_1^\gamma$ has attracted much attention 
in connection with the axial anomaly which is also 
relevant in the nucleon spin structure function $g_1^{p(n)}$. 
The polarized structure function $g_1^\gamma$ could be experimentally
studied in the polarized version of ep collider HERA,\cite{Barber,SVZ} 
or more directly measured by the polarized e$^+$e$^-$ collision in
the future linear collider. And the next-to-leading order QCD analysis
of $g_1^\gamma$ has been recently 
performed in the literature.\cite{SV,SU,GRS}

Now there exists another structure function $g_2^\gamma$ for the virtual
photon target, where the twist-3 effect is also relevant in addition to
the usual twist-2 effect. 
In this talk, we investigate the twist-3 effects in 
$g_2^\gamma$, to the leading order in QCD, and show that it is a sizable 
effect in contrast to the nucleon case,\cite{RLJ} where it appears 
to be small in the experimental data.\cite{E143,E155}

\section{Photon Structure Function $g_2^\gamma(x,Q^2,P^2)$ and Twist-3 Effects}

Let us consider the polarized deep inelastic scattering on a polarized virtual
photon target and study the virtual photon structure functions 
for the kinematical region:
\begin{eqnarray}
\Lambda^2 \ll P^2 \ll Q^2,
\end{eqnarray}
where $q^2=-Q^2$ ($p^2=-P^2$) is the mass squared of the probe (target) photon,
and $\Lambda$ is the QCD scale parameter.

The antisymmetric part of the structure tensor, $W^A_{\mu\nu\rho\tau}(p,q)$
for the target (probe) photon with momentum $p$ ($q$), 
relevant for the polarized structure, can be written in terms of the structure
functions, $g_1^\gamma$ and $g_2^\gamma$, as \cite{SU}
\begin{eqnarray}
W^A_{\mu\nu\rho\tau}&=&\frac{1}{(p\cdot q)^2}
\left[p\cdot q\,\epsilon_{\mu\nu\lambda\sigma}
{\epsilon_{\rho\tau}}^{\sigma\beta}q^\lambda p_\beta\ g_1^\gamma
\right.\nonumber\\
&-&\left.(\epsilon_{\mu\nu\lambda\sigma}
\epsilon_{\rho\tau\alpha\beta}q^\lambda p^\sigma q^\alpha
p^\beta-p\cdot q\,\epsilon_{\mu\nu\lambda\sigma}
{\epsilon_{\rho\tau}}^{\sigma\beta}q^\lambda p_\beta)\ g_2^\gamma\right] 
\end{eqnarray}
Here we note that we have the structure function $g_2^\gamma$
only for non-zero $P^2$, i.e. virtual photon. 

Now we decompose $g_2$ into twist-2 and twist-3 contributions:
\begin{eqnarray}
g_2=g_2^{\rm tw.2}+g_2^{\rm tw.3}.
\end{eqnarray}
where $g_2^{\rm tw.2}$ can be written by using
Wandzura-Wilczek relation:\cite{WW}
\begin{eqnarray}
g_2^{\rm tw.2}=g_2^{\rm WW}(x,Q^2)\equiv -g_1(x,Q^2)+\int_x^1
\frac{dy}{y}g_1(y,Q^2).
\end{eqnarray}

Experimental data for the nucleon show 
$g_2 \approx g_2^{\rm tw.2}=g_2^{\rm WW}$,
and now we ask what about the photon structure, especially virtual
photon case ?

\section{OPE and QCD Effects}
First we consider the operator product expansion (OPE) relevant for the 
photon structure functions. The OPE sandwitched between the photon states
can be decomposed into the twist-2 and twist-3 contributions as follows:
\begin{eqnarray}
&&\int d^4 x e^{iq\cdot x}
\langle\gamma(p,s)|J_{\mu}(x)J_{\nu}(0)^{[A]}|\gamma(p,s)\rangle \qquad
\nonumber\\
&&\sim \sum_{n} E_n^{(2)}(Q^2)
\langle\gamma(p,s)|R_n^{(2)}|\gamma(p,s)\rangle
+ \sum_{n} E_n^{(3)}(Q^2)
\langle\gamma(p,s)|R_n^{(3)}|\gamma(p,s)\rangle.\qquad
\end{eqnarray}
where $R_n^{(2)}$ and $R_n^{(3)}$ denote the twist-2 and twist-3
operators, respectively. For the nucleon the matrix element:
$\langle N(p,s)|R_n^{(3)}|N(p,s)\rangle$ is small.
Now we calculate the virtual 
photon structure functions 
arising from the so-called Box diagram, with $\langle e^4\rangle=
\sum_{i=1}^{N_f}e_i^4/N_f$, and $N_f$ being the number of active flavors.
It turns out that the twist-3 contribution to $g_2^\gamma$ is actually 
non-vanishing:
\begin{figure}[t]
\begin{center}
\vspace{-2.5cm}
\includegraphics[height=12cm]{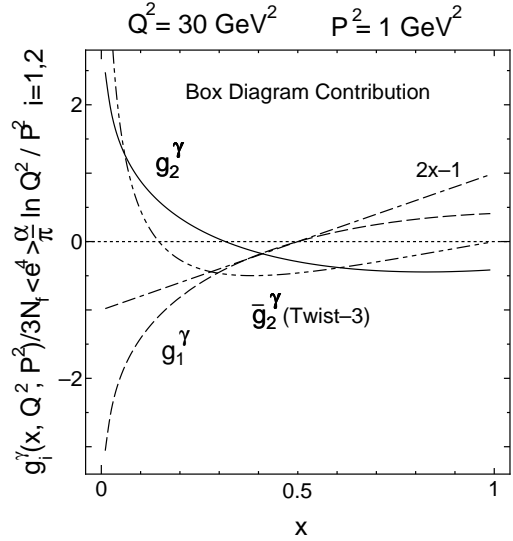}
\end{center}
\vspace{-2.5cm}
\caption{The Box-diagram contributions to $g_1^\gamma(x,Q^2,P^2)$ 
(dashed), $g_2^\gamma(x,Q^2,P^2)$ (solid) and 
${\bar g}_2^\gamma(x,Q^2,P^2)$ (dash-2dotted) for 
$Q^2=30$ GeV$^2$ and $P^2=1$ GeV$^2$ for $N_f=3$.}
\end{figure}
We have plotted the $g_1^\gamma$, $g_2^\gamma$ and ${\bar g}_2
^\gamma$ as a function of $x$ for the virtual photon target,
where $Q^2=30$ GeV$^2$ and $P^2=1$ GeV$^2$ for $N_f=3$ in Fig.1.

Now the $n$-th moment of the twist-3 part ${\bar g}_2^\gamma\equiv 
g_2^\gamma-g_2^{\gamma(WW)}$reads
\begin{eqnarray}
\int_0^1dx x^{n-1}{\bar g}_2^\gamma(x,Q^2,P^2)
&\approx& 
\frac{n-1}{2n}\vec{X}_n(Q^2/P^2,{\bar g}(Q^2),\alpha)
\vec{E}_n(1,{\bar g}(Q^2))
\end{eqnarray}
in the leading logarithmic order (LO) in QCD, where
\begin{eqnarray}
\vec{X}_n=\int_{{\bar g}}^gdg\frac{\vec{K}_n(g,\alpha)}{\beta(g)}
T\exp\left[\int_{{\bar g}}^gdg'\frac{{\hat\gamma}_n(g')}{\beta(g')}
\right]
\end{eqnarray}
with ${\hat \gamma}_n$ representing the mixing of the twist-3 
hadronic operators
and $\vec{K}_n$ describing the mixing between twist-3 hadronic and photonic 
operators.
Thus, the flavor non-singlet (NS) part reads
\begin{eqnarray}
\int_0^1 dx x^{n-1}{\bar g}_2^{\gamma(NS)}(x,Q^2,P^2)
&=&\frac{n-1}{n}\frac{\alpha}{4\pi}\cdot
\frac{1}{2\beta_0}
(-24N_f)(\langle e^4\rangle-\langle e^2\rangle^2)\frac{1}{n(n+1)}\nonumber\\
&&\times
\frac{1}{1+\lambda_{NS}^n/2\beta_0}
\frac{4\pi}{\alpha_s(Q^2)}\left\{
1-\left(\frac{\alpha_s(Q^2)}{\alpha_s(P^2)}\right)^{
\lambda_{NS}^n/2\beta_0+1}\right\}\nonumber
\end{eqnarray}
In the large $N_c$ limit ($N_c \rightarrow \infty$) we
have the 1-loop anomalous dimension of the NS twist-3 operator:\cite{ABH,KS3}
\begin{eqnarray}
&&\lambda_{NS}^n=\gamma_{(3)NS}^{0,n}=8C_F(S_n-\frac{1}{4}-\frac{1}{2n})
\qquad \mbox{where} \quad S_n=\sum_{j=1}^n\frac{1}{j}
\end{eqnarray}
up to $1/N_c^2$. We have plotted in Fig.2 the above result to LO in QCD for the
flavor non-singlet part. We expect the similar behavior for the flavor-singlet
contribution.

\begin{figure}[t]
\begin{center}
\vspace{-2.5cm}
\includegraphics[height=12cm]{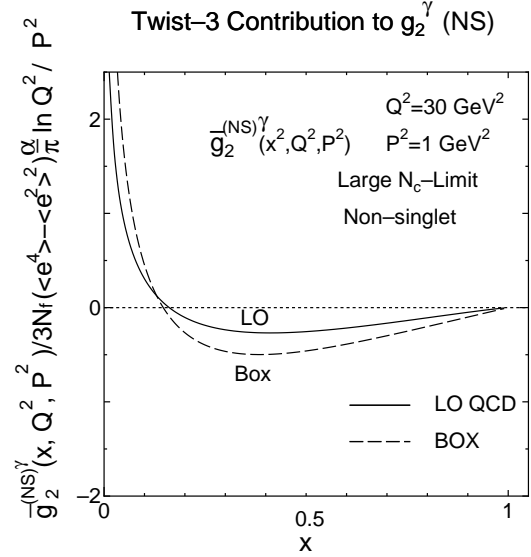}
\end{center}
\vspace{-2.5cm}
\caption{The Box-diagram (dashed) and the QCD LO (solid)
contributions to ${\bar g}_2^\gamma(x,Q^2,P^2)$  
for $Q^2=30$ GeV$^2$ and $P^2=1$ GeV$^2$ for $N_f=3$.}
\end{figure}

\section{Concluding Remarks}

We have investigated the twist-3 effect in $g_2^\gamma$ for the 
virtual photon target, in the LO QCD and have found it gives rise 
to a sizable effect.
More thorough QCD analysis including falvor-singlet part is now under
investigation.

\section*{Acknowledgements}
One of the authors (T.U.) 
thanks the organizers of the symposium, especially to 
Prof. Ma for the wonderful organization and the 
hospitality during the symposium.

\end{document}